\documentclass{article}
\usepackage{geometry}
\usepackage{amsfonts}
\usepackage{graphicx,color}
\usepackage{draftcopy}
\usepackage{color}   
\usepackage{hyperref}  
\usepackage[normalem]{ulem}
\usepackage{booktabs} 
\usepackage{multirow}
\usepackage{soul} 
\usepackage{microtype}

\definecolor{mygreen}{rgb}{0,0.5,0} 
\definecolor{myblue}{rgb}{0,0,1}
\definecolor{myred}{rgb}{1,0,0} 
\definecolor{mymagenta}{cmyk}{0,1,0,0.12}

\def\ba{\begin{eqnarray}}
\def\ea{\end{eqnarray}}

\def\lb{\label}

\usepackage{graphicx}

\usepackage{hyperref}
\hypersetup{colorlinks=true,linkcolor=blue,citecolor=blue}
\usepackage{cite}

\begin{document}

\title{\textbf{Photonic Discrete-Time Quantum Walks and Applications}}

\author{Leonardo Neves$^1$ and Graciana Puentes$^{2,3,}$\footnote{Correspondence: gpuentes@df.uba.ar}\, \vspace{1mm}
 \\
\textit{\normalsize{ $^1$Departamento de F\'isica, Universidade Federal de Minas Gerais, Belo Horizonte, MG 31270-901, Brazil}}  \\
\textit{\normalsize{ $^2$Departamento de F\'isica, Facultad de Ciencias Exactas y Naturales, Universidad de Buenos Aires,}}  \\ 
\textit{\normalsize{Ciudad Universitaria, 1428~Buenos Aires, Argentina}} \\
\textit{\normalsize{ $^3$CONICET-Universidad de Buenos Aires, Instituto de F\'{\i}sica de Buenos Aires (IFIBA),}} \\
\textit{\normalsize{Ciudad Universitaria, 1428~Buenos Aires, Argentina}}} 

\date{}

\maketitle

\begin{abstract}
We present a review of photonic implementations of discrete-time quantum walks (DTQW) in the spatial and temporal domains, based on spatial- and time-multiplexing techniques, respectively. Additionally, we propose a detailed novel scheme for photonic DTQW, using transverse spatial modes of single photons and programmable spatial light modulators (SLM) to manipulate them. Unlike all previous mode-multiplexed implementations, this scheme enables simulation of an arbitrary step of the walker, only limited, in principle, by the SLM resolution. We discuss current applications of such photonic DTQW architectures in quantum simulation of topological effects and the use of non-local coin operations based on two-photon hybrid entanglement.  \\

\noindent \textbf{Keywords} Quantum walks $\cdot$ Spatial-multiplexing $\cdot$ Time-multiplexing $\cdot$ Spatial light modulators
\end{abstract}

\section{Introduction}

The quantum walk is one of the most striking manifestations of how quantum interference leads to a strong departure between quantum and classical phenomena \cite{Kempe03}. In its discrete version, namely, the discrete-time quantum walk (DTQW) \cite{Kitagawa, Kitagawa2, Aharonov}, it offers a versatile platform for the exploration of a wide range of non-trivial geometric and topological phenomena both experimentally \cite{Kitagawa,Crespi,Alberti} and theoretically~\cite{Kitagawa2,Obuse,Shikano2,Asboth,Asboth2, Grunbaum,Wojcik,MoulierasJPB, Beenakker,Huber,Peano, Lu,Xia}. Furthermore, DTQWs are robust platforms for modeling a variety of dynamical processes from excitation transfer in spin chains \cite{Bose,Christandl} to energy transport in biological complexes~\cite{Plenio}. They enable to study multi-path quantum interference phenomena \cite{bosonsampling1,bosonsampling2,bosonsampling3,bosonsampling4} and can provide for a route to validation of quantum complexity \cite{validation1,validation2} and universal quantum computing~\cite{Childs}. Moreover,~multi-particle quantum walks warrant a powerful tool for encoding information in an exponentially larger space~\cite{PuentesPRA}, and for quantum simulations in biological, chemical and physical systems \cite{PuentesOL} in 1D and 2D geometries \cite{Peruzzo,OBrien,Silberhorn2D}. 

In this article, we present a review of photonic implementations of DTQW in the spatial \cite{Broome10} and temporal \cite{Silberhorn} domains, based on spatial- and time-multiplexing techniques, respectively \cite{PuentesCrystals17}. Additionally, we propose a detailed novel scheme for photonic DTQW, using transverse spatial modes of single photons and programmable spatial light modulators (SLM) to manipulate them. Unlike all previous mode-multiplexed implementations, this scheme enables simulation of an arbitrary step of the walker, only limited, in principle, by the SLM resolution. It works in an automated way by preparing the input state to the $n$-th step, applying a one-step evolution using the photon polarization as the quantum ``coin'', and, finally, measuring the probability distribution at the output spatial modes. We also discuss current applications of such photonic DTQW architectures in quantum simulation of topological effects and the use of non-local coin operations based on two-photon hybrid entanglement. Part of this review is based on the work by Puentes, selected as the cover story of a Special Issue on Quantum Topology, for the journal Crystals (MDPI)  in 2017 \cite{PuentesCrystals17}.


\section{Discrete-Time Quantum Walks}

 The basic step in the standard DTQW is given by a unitary evolution operator $U(\theta)=TR_{\vec{n}}(\theta)$, where $R_{\vec{n}}(\theta)$ is a rotation along an arbitrary direction $\vec{n}=(n_{x},n_{y},n_{z})$, given by 
$$
R_{\vec{n}}(\theta)=
\left( {\begin{array}{cc}
 \cos(\theta)-in_{z}\sin(\theta) & (in_{x}-n_{y})\sin(\theta)  \\
 (in_{x}+n_{y})\sin(\theta) & \cos(\theta) +in_{z}\sin(\theta)  \\
 \end{array} } \right) 
$$
   in the Pauli basis \cite{Pauli}. In this basis, the $y$-rotation is defined by an operator of the form  \cite{Pauli} 
$$
R_{y}(\theta)=
\left( {\begin{array}{cc}
 \cos(\theta) & -\sin(\theta)  \\
 \sin(\theta) & \cos(\theta)  \\
 \end{array} } \right). 
$$
{This is the so-called coin operator as it will act on a two-dimensional degree of freedom of a given quantum system playing the role of the quantum coin in the DTQW protocol.} We~note that the coin operation acts on polarization in the case of photons, or on spin in the case of atoms. The above rotation is followed by a spin- or polarization-dependent translation $T$ given by 
$$
T=\sum_{x}|x+1\rangle\langle x | \otimes|H\rangle \langle H| +|x-1\rangle \langle x| \otimes |V\rangle \langle V|,
$$
 where $H=(1,0)^{T}$ and $V=(0,1)^{T}$.
The evolution operator for a discrete-time step is equivalent to that generated by a Hamiltonian $H(\theta)$, such that $U(\theta)=e^{-iH(\theta)}$ ($\hbar=1$), with 
$$
H(\theta)=\int_{-\pi}^{\pi} dk[E_{\theta}(k)\vec{n}(k).\vec{\sigma}] \otimes |k \rangle \langle k|
 $$
and $\vec{\sigma}$ the Pauli matrices, which readily reveals the {spin-orbit or polarization-spatial coupling mechanism in the atomic or photonic system, respectively}. The quantum walk described by $U(\theta)$ has been realized experimentally in several \mbox{systems~\cite{photons,Silberhorn,Broome10,ions,coldatoms}} and has been shown to posses chiral symmetry and display Dirac-like dispersion relation given by $\cos(E_{\theta}(k))=\cos(k)\cos(\theta)$. We note that the spectrum of the system will depend on the choice of branch cut. We fix the branch cut at the quasi-energy gap \cite{energy_shift1,energy_shift2, PuentesCrystals17}.

\section{Photonic Implementations}
\vspace{-6pt}
\subsection{Spatial Multiplexed Discrete-time Quantum Walk}
\label{sec:spa_multiplex}

The experimental scheme for photonic implementation of DTQW via spatial multiplexing was first introduced by Broome et al.\ \cite{Broome10} {using calcite beam displacers, and later on revisited by Sansoni~et~al.~\cite{Sansoni12} using integrated multi-mode interferometers.} The Hilbert space for the DTQW is determined by $2n+1$ multiplexed longitudinal spatial modes of a single photon coupled to a coin encoded in its two dimensional polarization subspace $\{|H\rangle, |V\rangle\}$. These spatial modes $\{|j\rangle\}$ are labeled as  $j=\pm(n-2k)$ with $k=0,1,\ldots,\lfloor n/2\rfloor$, where $n$ denotes the current walker's step. Single-photons  created via
spontaneous parametric down-conversion in a PPKTP crystal are injected into a free-space reference mode $|j\rangle=|0 \rangle$. This  mode is spatially multiplexed by a sequence of calcite polarizing beam-displacers (CBD). Arbitrary initial coin polarization states are prepared by a polarizing beam-splitter and a combination of half- (HWP) and quarter waveplates. Subsequently, a combination of a HWP and a CBD implements the one-step evolution. By concatenating $n$ of such arrangements one can implement $n$ steps of a quantum walk  (see Figure~\ref{fig:setup}a).  
Coincident detection of photons at avalanche detectors (APDs) (4.4 ns time window) herald a successful run of the walk. The typical number steps implemented with spatial-multiplexed schemes is of order $n \approx 10$ \cite{Broome10} (see \ref{fig:setup}a).{  On the other hand, the integrated waveguide approach is based on an integrated waveguide architecture which enables concentration of a large number of optical
elements on a small chip, achieving phase
stability due to the monolithic structure. In the waveguide architecture calcite beam displacers (CBD) are replaced by directional couplers
(DCs) i.e., structures in which two waveguides, brought
close together for a certain interaction length, couple by
evanescent field \cite{Sansoni12} (see \ref{fig:setup}b)}. 

\subsection{Time Multiplexed Discrete-Time Quantum Walk}

\begin{figure}[h!]
\centering
\includegraphics[width=0.8\textwidth]{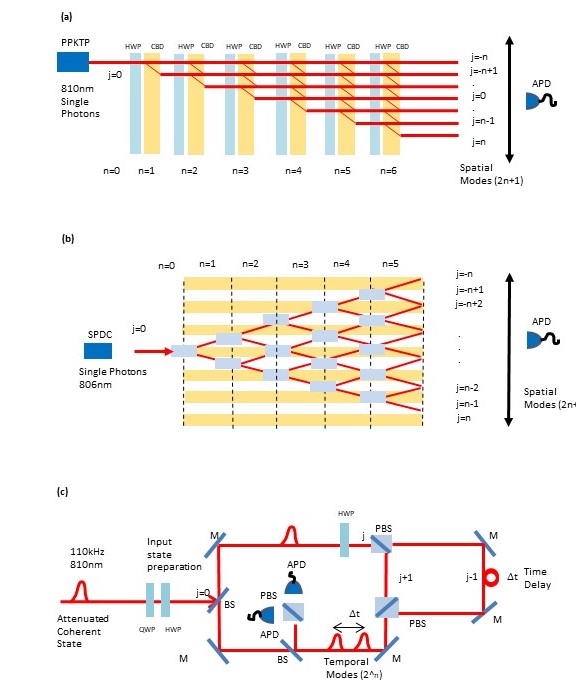}
\caption{\label{fig:setup} Experimental scheme for implementation of DTQWs  via:  ({\bf a}) spatial multiplexing using calcite beam displacers, ({\bf b})  spatial multiplexing using integrated waveguides, ({\bf c})  time multiplexing using a calibrated delay line (see text for details). }
\end{figure}

The experimental scheme for DTQW via time-multiplexing was first introduced in Ref.~\cite{Silberhorn}. The~Hilbert space for the DTQW is determined by a single spatial mode $|j\rangle=| 0 \rangle$ and $2^n$  multiplexed temporal modes $|k \rangle$ (for $k=1,2,...,2^n$), with $n$ the step number, coupled to a coin operator in a two dimensional polarization subspace $(|H\rangle, |V\rangle)$ (see Figure~\ref{fig:setup}b). Equivalent single-photon states (on average) are generated with an attenuated pulsed diode laser centered at 810 nm and with 111 kHz repetition rate (RR).  The initial state of the photons is controlled via half-wave plates (HWPs) and quarter-wave plates (QWPs), in order to produce eigenstates of chirality $|\psi_{0}^{\pm}\rangle=|0\rangle \otimes (|H\rangle \pm i|V\rangle)/\sqrt{2}$. Inside the loop, the rotation ($R_{n}(\theta)$) is implemented by an HWP with its optical axis oriented at an angle~$\theta/2$. 
The~spin-dependent translation is realized in the time domain via a polarizing beam splitter (PBS) in combination with a calibrated fiber delay line, in which horizontally polarized light follows a longer path {and is delayed by an amount $\Delta t$}. The resulting temporal difference  between both polarization components corresponds to a step in the spatial domain  ($x \pm 1$). Polarization controllers (PCs) are introduced to compensate for arbitrary polarization rotations in the fibers. After implementing the time-delay, the time-bins are interferometrically recombined in a single spatial mode by means of a second PBS and are re-routed into the fiber loops by means of silver mirrors. The~detection is realized by coupling the photons out of the loop by a {beam splitter} (BS) with a probability of 5$\%$ per step.~Avalanche photodiodes (APDs) are employed to measure the photon arrival time and polarization properties. The probability that a photon undergoes a full round-trip is given by the overall coupling efficiency  ($>$70\%) and the overall losses in the setup resulting in $\eta= 0.50$.  The~average photon number per pulse is controlled via neutral density filters and is below $\langle n \rangle <0.003$ to ensure negligible contribution from multi-photon events. This scheme allows for implementing a large number of steps ($n \approx 20$) in a compact architecture, thus improving upon spatially multiplexed architectures (see~Figure~\ref{fig:setup}c).

\subsection{Discrete-Time Quantum Walk in Transverse Propagation Modes using Spatial Light Modulators}   \label{sec:QW_SLM} 

\begin{figure}[t]
\centering
\includegraphics[width=0.7\columnwidth]{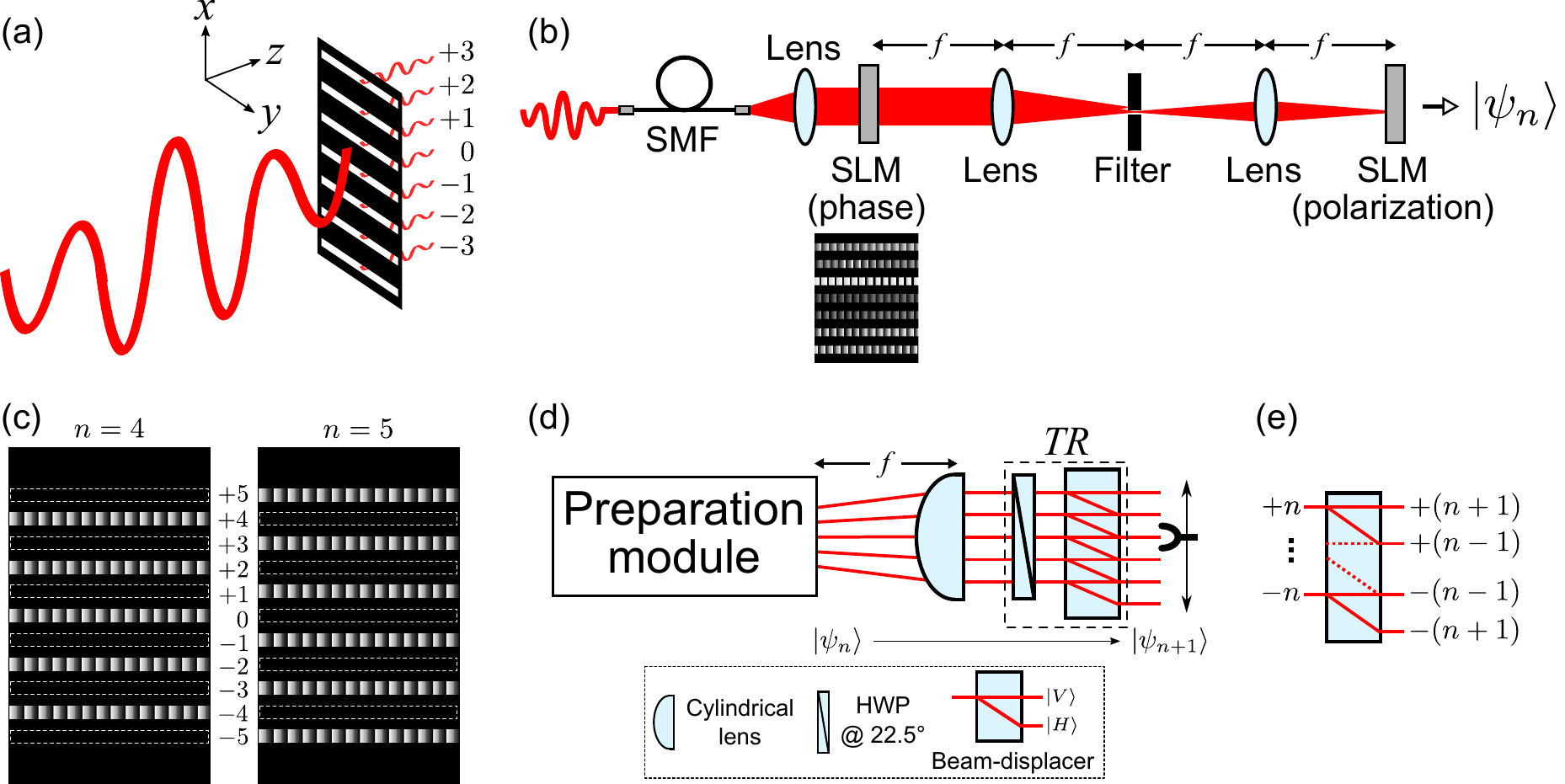}
\caption{\label{fig:SLM} ({\bf a}) Discretization of a single-photon spatial amplitude profile in transverse modes along the $x$-direction. ({\bf b}) Sketch of the proposed optical setup for preparing the $n$-th step walker-coin state~((\ref{eq:input})) encoded in the transverse modes and polarization of a single-photon field: SMF, single mode fiber for spatial filtering; SLM, programmable spatial light modulator (see text for details). ({\bf c}) Phase masks addressed at the phase-only SLM for preparing the state given by Equation~(\ref{eq:chi_n}) with $n=4$ and 5. The dashed rectangles indicate empty transverse modes. ({\bf d}) Optical module that implements one step ($|\psi_n\rangle\rightarrow|\psi_{n+1}\rangle$) of the 1D DTQW proposed here (see text for details). ({\bf e}) Numbering convention of the spatial modes exiting the beam-displacer \cite{Broome10}. } 
\end{figure}

In this section, let us consider the sites of the 1D DTQW represented by transverse spatial modes of a single photon. To be more specific, if the photon propagates in the $z$-direction, these sites will be defined in one dimension, say $x$, of its transverse propagation plane. The walker's Hilbert space will be spanned by the basis $\{|j\rangle: j\in\mathbb{Z}\}$, where the state $|0\rangle$ corresponds to the spatial mode in the optical axis of the system and $\{|j>0\rangle\}$ ($\{|j<0\rangle\}$) to the upper (lower) modes, as illustrated in~Figure~\ref{fig:SLM}a. {Instead of the approach of Ref.~\cite{Broome10} described in Section~\ref{sec:spa_multiplex}, we shall follow the one} proposed and demonstrated (for one step) by Francisco et al. in an ingenious but intricate setup \cite{Francisco06}, mainly regarding the encoding of the quantum coin in the upper and lower regions of the $x$-axis. 
Here, we propose a simpler and more intuitive approach: the quantum coin will be encoded in the polarization of the photon in the horizontal/vertical basis, i.e., $\{|H\rangle,|V\rangle\}$. Therefore, the conditional translation operator will be written as
\begin{equation}  \label{eq:T}
T=\sum_j|j+1\rangle\langle j|\otimes|H\rangle\langle H|+|j-1\rangle\langle j|\otimes|V\rangle\langle V|.
\end{equation}
Assuming an unbiased coin operator
\begin{equation}   \label{eq:R}
R=\frac{1}{\sqrt{2}}\left(\begin{array}{cc}
1 & 1 \\
1 & -1
\end{array}\right),
\end{equation}
and starting in the initial walker-coin state $|\psi_0\rangle=|0\rangle|H\rangle$, after $n$ steps it will evolve into
\begin{equation}  \label{eq:input}
|\psi_n\rangle=(TR)^n|\psi_0\rangle=\frac{1}{\sqrt{n+1}} \sum_{j=0}^{n}e^{i\phi_{n-2j}}|n-2j\rangle|\theta_{n-2j}\rangle,
\end{equation}
where $\phi_{n-2j}=0$ or $\pi$, and $|\theta_{n-2j}\rangle=\cos(\theta_{n-2j})|H\rangle+\sin(\theta_{n-2j})|V\rangle$ is the linear polarization state of the coin in the $(n-2j)$-th walker's spatial mode. For instance, if $n=4$
\begin{equation}  \label{eq:psi4}
|\psi_4\rangle\propto|+4\rangle|H\rangle+|+2\rangle\left(\frac{3|H\rangle+|V\rangle}{\sqrt{10}}\right)+
|0\rangle\left(\frac{|V\rangle-|H\rangle}{\sqrt{2}}\right)-
|-2\rangle\left(\frac{|V\rangle-|H\rangle}{\sqrt{2}}\right)-|-4\rangle|V\rangle.
\end{equation}
The corresponding probability distribution of walker's position after $n$ steps will be computed by
\begin{equation}   \label{eq:P_n}
P_n(j)=|(\langle H|\langle j|)|\psi_n\rangle|^2+|(\langle V|\langle j|)|\psi_n\rangle|^2.
\end{equation}

To study the 1D DTQW within the scenario described above, we shall propose a feasible optical setup which is divided in two modules. The first is designed to prepare the state given by  Equation~(\ref{eq:input}) for an arbitrary value of $n$, only limited, in principle, by the resolution of the devices used to manipulate the photonic transverse spatial modes, as will be discussed below. The second will implement one step of the protocol, namely, the unitary operation $TR$, with $T$ and $R$ given by Equations~(\ref{eq:T}) and (\ref{eq:R}), respectively. With the preparation module, the probability distributions given by Equation~(\ref{eq:P_n}) can be directly measured. In addition, by concatenating it with the one-step module, it will be possible to implement one step of the quantum walk from $n$ to $n+1$. In both cases, one will be able to simulate the 1D DTQW for steps (large values of $n$) that cannot be achieved with implementations like time- or spatial-multiplexing. Next, we describe the proposed optical modules and the measurement stage from which one estimates $P_n(j)$.

\subsubsection{Preparation Module}

Figure~\ref{fig:SLM}b sketches the optical module we propose to prepare the walker-coin state in the $n$-th step of a quantum walk (Equation~(\ref{eq:input})) using the transverse spatial modes and polarization of single photons. It is divided in two sections: the first, designed to prepare the spatial part of the state, and~the second, to couple it with the polarization degree of freedom. A central feature for the operation of this module will be the use of programmable spatial light modulators (SLMs). These devices, based on liquid crystal display (LCD), consist of a two-dimensional array of pixels, each of which, when properly configured, can control the amplitude, phase or polarization of the incident light field \cite{SLMBook}. Recently, they have been used in a variety of quantum information protocols \cite{Puentes04,Marques15,Prosser17,Bouchard17}. 

Let
$\int\!d\vec{r}\,\psi(\vec{r})|1\vec{r}\rangle\otimes|H\rangle$ be the
quantum state of a paraxial and monochromatic single-photon multimode field horizontally polarized, where $\vec{r}=(x,y)$ is the transverse position coordinate and $\psi(\vec{r})$ is the normalized transverse probability
amplitude. It can be generated, for instance, from a heralded parametric down-conversion source. By manipulating the transverse amplitude $\psi(\vec{r})$ with the technique developed by Prosser et al.\ in \cite{Prosser13}, it is possible to prepare arbitrary superpositions of the form $\sum_j\beta_j|j\rangle$ with $\sum_j|\beta_j|^2=1$, where $\{|j\rangle\}$ represent the orthogonal transverse modes in the $x$-direction. In short, the technique works as follows. Consider an SLM that modulates only the phase of the incident profile $\psi(\vec{r})$,  which, for simplicity, is assumed to be constant across the region of modulation. A phase mask with an array of $d$ rectangular regions, each one filled with a blazed diffraction grating, is displayed on the LCD screen (an example of mask with $d=7$ is shown in the inset of Figure~\ref{fig:SLM}b). The photon profile has its phase modulated by this mask and, in the far field (the~back focal plane of the lens), it is diffracted into orders ($0,\pm 1,\ldots$) when it reaches regions with gratings; otherwise, it goes to the zeroth order. If one chooses the $+1$ order to prepare the states, the moduli of its complex coefficients will be adjusted according to the phase modulation depth of each grating, which defines the intensity diffracted to that order. On the other hand, the phases of these coefficients are adjusted with a constant phase value added to the gratings. Finally, the $+1$ diffraction order is filtered from the others by a slit diaphragm and the emerging photon will be in a coherent superposition of the $d$ transverse ``which-slit'' modes $\{|j\rangle\}$. In particular, one can prepare states given~by
\begin{equation}  \label{eq:chi_n} 
|\chi_n\rangle=\left(\frac{1}{\sqrt{n+1}}\sum_{j=0}^{n}e^{i\phi_{n-2j}}|n-2j\rangle\right)\otimes|H\rangle,
\end{equation}
where $\phi_{n-2j}=0$ or $\pi$, and $n$ is a nonnegative integer. For a given  $n$, one addresses a phase mask to the SLM with $d=n+1$ slit/diffraction gratings symmetrically distributed from the highest modes $j=\pm n$. Figure~\ref{fig:SLM}c shows an example of such mask for $n=4$ and $n=5$. As a technical note, since the states we want to prepare are uniform (see Equations~(\ref{eq:input}) and (\ref{eq:chi_n})), the phase modulation depth of all gratings displayed at the SLM will be equal. Thus, by setting it to $2\pi$, one achieves, in theory, 100\% of efficiency in $+1$ diffraction order. 

To generate the state given by Equation~(\ref{eq:input}) from the state of Equation~(\ref{eq:chi_n}), one must implement polarization rotations conditioned to the transverse mode positions, which is represented by the unitary operator
\begin{equation}    \label{eq:cond_rot}
\sum_{j=0}^{n}|n-2j\rangle\langle n-2j|\otimes \mathcal{R}(\vartheta_{n-2j}),
\end{equation}
where
\begin{equation}  \label{eq:rot}
\mathcal{R}(\vartheta)=\left(\begin{array}{cc}
\cos\vartheta & -\sin\vartheta \\
\sin\vartheta & \cos\vartheta
\end{array}\right)
\end{equation}
will transform $|H\rangle$ into an arbitrary state of linear polarization (it should not be confused with the coin operation). By applying this transformation on $|\chi_n\rangle$ with the appropriate $\mathcal{R}(\vartheta_{n-2j})$'s, one generates the desired state $|\psi_n\rangle$. 

These spatially dependent polarization rotations can be implemented by an SLM properly configured for the task \cite{Moreno03}. There exist different techniques for different types of SLMs which enable each pixel of the device to work effectively as controllable polarization rotator \cite{Davis00,Moreno07}. The details of this process are beyond the scope of the present work; we are just interested in its useful effects. With this SLM, the transformation (\ref{eq:cond_rot}) onto the state (\ref{eq:chi_n}) can be implemented by imaging the transverse spatial modes of $|\chi_n\rangle$ on the LCD screen and applying the proper modulation on the input polarization $|H\rangle$. As shown in Figure~\ref{fig:SLM}b, this is achieved with a $4f$ lens system that will image the filtered output field of the phase-only SLM (the state $|\chi_n\rangle$) onto a polarization-rotator SLM. 

This concludes the operation of the proposed optical module for preparing a walker-coin state in the $n$-th step of a 1D DTQW, encoded in the transverse spatial modes and polarization, respectively, of a single photon. As mentioned earlier, the maximum value of $n$ will be limited, in this case, by the resolution of SLM. To illustrate this, assume that a phase-only SLM has $2N$ pixels in the direction where the transverse modes are encoded. If each mode is encoded in a row and separated by another row of pixels, both with one-pixel width, it would be possible to define $N$ distinguishable modes. In turn, this would enable us, in principle, to prepare the walker-coin state ((\ref{eq:input})) up to $n=\lfloor N/2\rfloor$. For~instance, if $2N=1920$ \cite{SLMBook}, then $n=480$, which represents much larger steps that can be achieved with other approaches. 

After state preparation, one can determine the probability distribution (\ref{eq:P_n}) by recording the photon count rates at each of the $2n+1$ output transverse modes of the second SLM (see Figure~\ref{fig:SLM}b) and normalizing them at each mode to the total number of counts. This can be achieved either with an array of $2n+1$ single-photon detectors or with one single-photon detector scanning the corresponding transverse modes. The detection system must be placed right after the second SLM, to avoid the modes to diffract and interfere. Alternatively, as described below, one can made a transverse-to-longitudinal mode conversion and put the detector at greater distances from the preparation module. 

\subsubsection{One-step Module}

With the quantum coin encoded in the photon polarization states $|H\rangle$ and $|V\rangle$, the coin operation $R$, given by Equation~(\ref{eq:R}), is straightforwardly implemented by a half wave plate (HWP) set to $22.5^\circ$. On the other hand, to implement the conditional translation $T$, given by (Equation~(\ref{eq:T})), one is naturally led to think in using some birefringent material. However, this element, alone, will not be useful if the transverse modes are allowed to propagate in continuous free space, as it will not avoid these modes to diffract and eventually interfere, making difficult, if not impossible, to characterize the walker's translation. To keep the discrete nature of the protocol while working with a degree of freedom that is discretized in the plane of state preparation but will evolve in a continuous variables space, one must enforce the discretization along all propagation planes. Here, this can be achieved by placing a cylindrical lens with focal length $f$ at a distance $f$ from the second SLM of the preparation module, as shown in Figure~\ref{fig:SLM}d. In doing so, the transverse modes from the output preparation plane will be converted into longitudinal modes along the remaining propagation planes. With this conversion one can simply use a birefringent calcite beam-displacer to implement $T$. This optical element may be cut to directly transmit vertically polarized light and induce a lateral displacement on the horizontally polarized light into the neighboring mode, as illustrated in the inset of Figure~\ref{fig:SLM}d. To achieve the desired interference effect, the input and displaced modes must be matched, and this can be done by an appropriate combination between the lens focal length and the lateral displacement provided by the calcite. 

Therefore, the one-step module $TR$ for an input state $|\psi_n\rangle$ given by (Equation (\ref{eq:input})) is comprised by a cylindrical lens to implement the transverse-to-longitudinal mode conversion, and a HWP at $22.5^\circ$ and a beam-displacer that implement $R$ and $T$, respectively. After this step, one can measure the probability distribution $P_{n+1}(j)$ (Equation~(\ref{eq:P_n})) as described earlier. The entire process is sketched in Figure~\ref{fig:SLM}d. Figure~\ref{fig:SLM}e shows the convention adopted for numbering spatial modes exiting the beam-displacer. Such an arrangement for coin operation and conditional translation has been proposed and successfully demonstrated by Broome et al.\ for 1D DTQW up to the sixth step \cite{Broome10}.

\section{Applications}

In this section, we propose different applications amenable to each type of DTQW implementation, namely via spatial-mode multiplexing, temporal-mode multiplexing, or using spatial light modulators~(SLMs). 

\subsection{Applications via Spatial Multiplexed DTQW: Split-Step Quantum Walk}

We present two different examples of non-trivial geometrical Zak phase structure in the holonomic sense, i.e.,the  equivalent to the Berry phase across the Brillouin zone. The first DTQW protocol consists of two consecutive spin-dependent translations $T$ and rotations $R$, such that the unitary step becomes \mbox{$U(\theta_1,\theta_2)=TR(\theta_1)TR(\theta_2)$}, as described in detail in \cite{Kitagawa2}. The so-called ``split-step'' quantum walk has been shown to possess a non-trivial topological landscape given by topological sectors, which are delimited by continuous 1D topological boundaries. These topological sectors are characterized by topological invariants, such as the winding number, taking integer values $W=0,1$.
  The dispersion relation for the split-step quantum walk results in \cite{Kitagawa2}:
$$
\cos(E_{\theta,\phi}(k))=\cos(k)\cos(\theta_1)\cos(\theta_2)-\sin(\theta_1)\sin(\theta_2).
$$

The 3D-norm for decomposing the quantum walk Hamiltonian of the system in terms of Pauli matrices $H_{\mathrm{QW}}=E(k)\vec{n} \cdot \vec{\sigma}$  becomes \cite{Kitagawa}: \\
\begin{equation}
\begin{array}{ccc}
n_{\theta_1,\theta_2}^{x}(k)&=&\displaystyle\frac{\sin(k)\sin(\theta_1)\cos(\theta_2)}{\sin(E_{\theta_1,\theta_2}(k))}\vspace{3pt},\\[5mm]
n_{\theta_1,\theta_2}^{y}(k)&=&\displaystyle\frac{\cos(k)\sin(\theta_1)\cos(\theta_2)+\sin(\theta_2)\cos(\theta_1)}{\sin(E_{\theta_1,\theta_2}(k))}\vspace{3pt},\\[5mm]
n_{\theta_1,\theta_2}^{z}(k)&=&\displaystyle\frac{-\sin(k)\cos(\theta_2)\cos(\theta_1)}{\sin(E_{\theta_1,\theta_2}(k))}.\\
\end{array}
 \end{equation}
 
The dispersion relation and topological landscape for the split-step quantum walk was analyzed in detail in \cite{Kitagawa2}. We now turn to our second example.

\subsection{Applications via Temporal Multiplexed DTQW: Quantum Walk with Non-Commuting Rotations}

The second example consists of two consecutive non-commuting rotations in the unitary step of the DTQW \cite{PuentesCrystals17}. The second rotation along the $x$-direction by an angle $\phi$, such that  the unitary step becomes $U(\theta,\phi)=TR_{x}(\phi)R_{y}(\theta)$, where $R_{x}(\phi)$ is given in the same basis \cite{Pauli} by:\vspace{12pt}
$$
R_{x}(\phi)=
\left( {\begin{array}{cc}
 \cos(\phi) & i\sin(\phi)  \\[2mm]
i \sin(\phi) & \cos(\phi)  \\
 \end{array} } \right).
$$
 
   The modified dispersion relation becomes:
\begin{equation}
\cos(E_{\theta,\phi}(k))=\cos(k)\cos(\theta)\cos(\phi)+\sin(k)\sin(\theta)\sin(\phi),
\end{equation}
 where we recover the Dirac-like dispersion relation for $\phi=0$, as expected.
The 3D-norm for decomposing the Hamiltonian of the system in terms of Pauli matrices  becomes: 
\begin{equation}
\begin{array}{ccc}
n_{\theta,\phi}^{x}(k)&=&\displaystyle\frac{-\cos(k)\sin(\phi)\cos(\theta)+\sin(k)\sin(\theta)\cos(\phi)}{\sin(E_{\theta,\phi}(k))}\vspace{3pt},\\[5mm]
n_{\theta,\phi}^{y}(k)&=&\displaystyle\frac{\cos(k)\sin(\theta)\cos(\phi)+\sin(k)\sin(\phi)\cos(\theta)}{\sin(E_{\theta,\phi}(k))}\vspace{3pt},\\[5mm]
n_{\theta,\phi}^{z}(k)&=&\displaystyle\frac{-\sin(k)\cos(\theta)\cos(\phi)+\cos(k)\sin(\theta)\sin(\phi)}{\sin(E_{\theta,\phi}(k))}.\\
\end{array}
 \end{equation}

\begin{figure}[t]
\label{fig:3}
\centering
\includegraphics[width=0.55\linewidth]{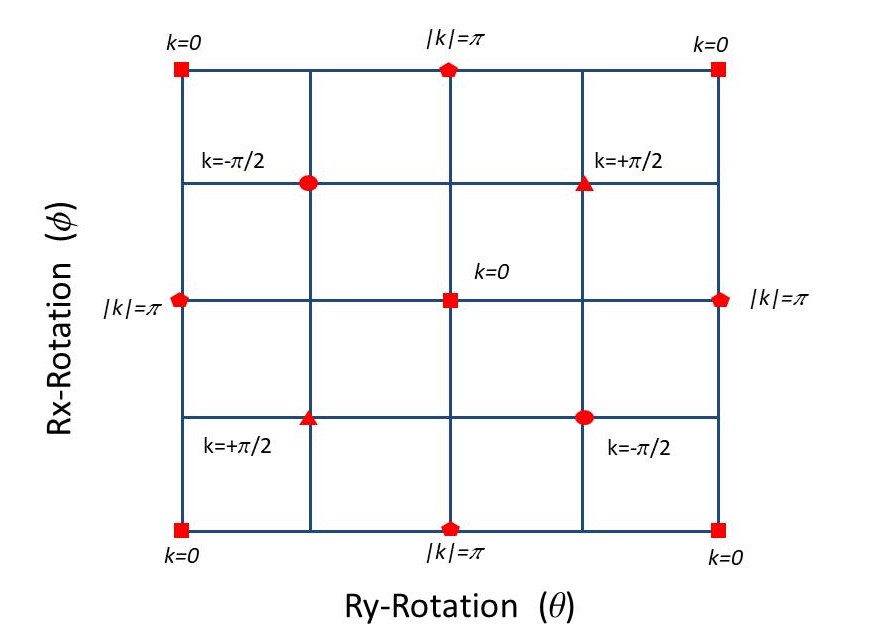} 
\caption{Non-trivial phase diagram for the quantum walk with consecutive non-commuting rotations, indicating gapless Dirac points where quasi-energy gap closes for different values of quasi-momentum: squares ($k=0$), pentagons ($|k|=\pi$), romboids ($k=+\pi/2$), and circles ($k=-\pi/2$). These discrete Dirac points represent topological boundaries of dimension zero. They endow the system with a non-trivial~topology  \cite{PuentesCrystals17}. 
\label{f1}}
\end{figure}

As anticipated, this system has a non-trivial phase diagram with a larger number of gapless points for different momenta as compared to the system consisting of a single rotation. Each of these gapless points represent topological boundaries of dimension zero, where topological invariants are not defined. Unlike the ``split-step'' quantum walk described previously, this system does not contain continuous topological boundaries. We calculated analytically the gapless Dirac points and zero-dimension topological boundaries for the system. Using basic trigonometric considerations, it can be shown that the quasi-energy gap closes at 13 discrete points for different values of quasi-momentum $k$. The phase diagram indicating the Dirac points where the gap closes for different momentum values is shown in Figure \ref{f1}. Squares correspond to Dirac points for $k=0$, circles correspond to Dirac points for $k=-\pi/2$, romboids correspond to Dirac points for $k=+\pi/2$, and pentagons correspond to Dirac points for $|k|=\pi$.  This geometric  structure in itself is novel and topologically non-trivial. Moreover, it~has not been studied in detail before.

\section{Geometric Zak Phase Calculation}

Phases arising during the quantum evolution of a particle can have different origins. A type of geometric phase, the so-called Berry phase \cite{Berry}, can be ascribed to quantum particles which return adiabatically to their initial state, but remember the path they took by storing this information on a geometric phase ($\Phi$), defined as \cite{Berry, Hannay}:
\begin{equation}
e^{i\Phi}=\langle \psi_{\mathrm{ini}}|\psi_{\mathrm{final}} \rangle.
 \end{equation}

 Geometric phases carry several implications: they modify material properties of solids, such as conductivity in Graphene 
 \cite{Berrygraphene,Delplace}, they are responsible for the emergence of surface edge-states in topological insulators, whose surface electrons experience a geometric phase \cite{Berrytopoinsul, Berrytopoinsul1, Berrytopoinsul2}, they can modify the outcome of molecular chemical reactions  \cite{Berrychemestry}, and  could even have implications for quantum information technology, via the  Majorana particle \cite{Berrymayorana}, or can bear close analogies to gauge field theories and differential geometry \cite{BerryGauge}.

We will now give expressions for the Zak Phase, i.e., the geometric phase acquired due to quantum evolution across the Brillouin Zone, in two different scenarios. These scenarios are casted by the following generic Hamiltonian
\begin{equation}
H\sim n_x \sigma_x+n_y \sigma_y+ n_z \sigma_z.
\end{equation}

The Hamiltonian to be described differ by a multiplying factor and by the expression of the $n_i$. However,~since the eigenvectors are the only quantities of interest for the present problem, the overall constants of this Hamiltonian can be safely ignored.

Now, our generic Hamiltonian is given by the~matrix
\begin{equation}
H=
\left(
\begin{array}{cc}
  n_z \qquad    n_x-in_y\\[2mm]
 n_x+i n_y \qquad   -n_z  
 \end{array} 
\right)
\end{equation}
and has the following eigenvalues
\begin{equation}
\lambda=\pm \sqrt{n_x^2+n_y^2+n_z^2}.
\end{equation}

The normalized eigenvectors then result as
\begin{equation}
|V_\pm\rangle= 
\left(
\begin{array}{cc}
\displaystyle  \frac{n_x+i n_y}{\sqrt{2n_x^2+ 2n_y^2+2n_z^2\mp 2n_z\sqrt{n_x^2+n_y^2+n_z^2}}}    \\[8mm]
\displaystyle  \frac{n_z\mp \sqrt{n_x^2+n_y^2+n_z^2}}{\sqrt{2n_x^2+ 2n_y^2+2n_z^2\mp 2n_z\sqrt{n_x^2+n_y^2+n_z^2}}}   
\end{array}
\right).
\end{equation}

Please note that the scaling $n_i\to\lambda n_i$ does not affect the result, which is as it should be. This follows from the fact that two Hamiltonians
related by a constant have the same eigenvectors.

The Zak phase ($\Phi_{Zak}=Z$) for each band ($\pm$) can be expressed as: 
\begin{equation}
Z_{\pm}=i\int_{-\pi/2}^{\pi/2} dk \langle V_{\pm}|\partial_k V_{\pm}\rangle.
\end{equation}

We will now apply these concepts to some specific examples.

\subsection{Split-Step Quantum Walk}

We first consider the split-step quantum walk \cite{Kitagawa2, PuentesJPB16}. This corresponds to a quantum walk with unitary step give by $U(\theta_1, \theta_2)=TR(\theta_1)TR(\theta_2)$, which can be readily implemented via spatial multiplexed DTQW as proposed in \cite{Kitagawa2, PuentesJPB16}. In this example, the normals $n_i$ are of the following form:

\begin{equation}
\begin{array}{ccc}
n_{\theta_1,\theta_2}^{x}(k)&=&\displaystyle\frac{\sin(k)\sin(\theta_1)\cos(\theta_2)}{\sin(E_{\theta_1,\theta_2}(k))}\vspace{3pt},\\[5mm]
n_{\theta_1,\theta_2}^{y}(k)&=&\displaystyle\frac{\cos(k)\sin(\theta_1)\cos(\theta_2)+\sin(\theta_2)\cos(\theta_1)}{\sin(E_{\theta_1,\theta_2}(k))}\vspace{3pt},\\[5mm]
n_{\theta_1,\theta_2}^{z}(k)&=&\displaystyle\frac{-\sin(k)\cos(\theta_2)\cos(\theta_1)}{\sin(E_{\theta_1,\theta_2}(k))}.\\
\end{array}
 \end{equation}

We consider the particular case that $n_z=0$. By taking one of the angle parameters such that $n_z=0$, it follows that the eigenvectors of the Hamiltonian are:
\begin{equation}
|V_\pm\rangle= 
\frac{1}{\sqrt{2}}\left(
\begin{array}{cc}
  e^{-i\phi(k)}   \\
  \mp 1 
\end{array}
\right),\qquad 
\tan\phi(k)=\frac{n_y}{n_x}.
\end{equation}

There are two choices for $n_z=0$, which are $\theta_1=0$ or $\theta_2=0$.
The Zak phase for each band takes the same value and results in  \cite{PuentesCrystals17}:
\begin{equation}
Z=Z_{\pm}=i\int_{-\pi/2}^{\pi/2}dk \langle V_{\pm}|\partial_{k} V_{\pm}\rangle,
\end{equation}
\begin{equation}
Z=i\int_{-\pi/2}^{\pi/2} dk \langle V_{\pm}|\partial_k V_{\pm}\rangle=\phi(-\pi/2)-\phi(\pi/2),
\end{equation}
from where it follows that
\begin{equation}
Z=\frac{\tan(\theta_2)}{\tan(\theta_1)}.
\end{equation}

\begin{figure}
\label{fig:2}
\centering
\includegraphics[width=0.8\linewidth]{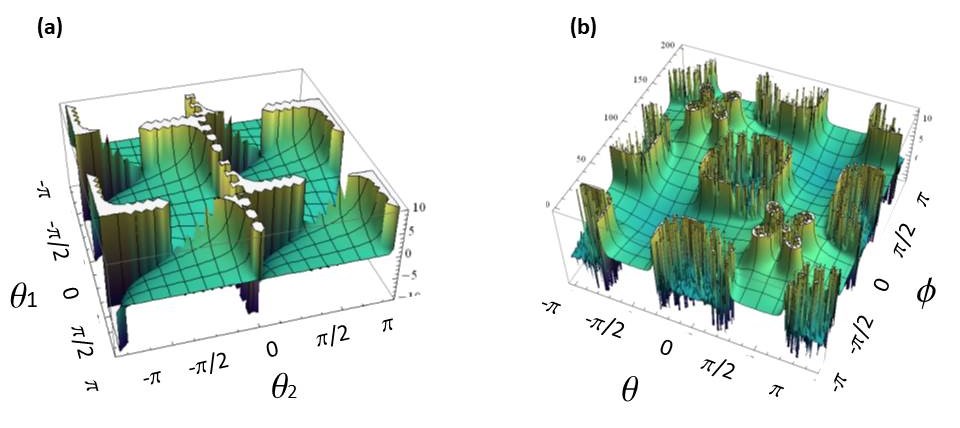} \caption{(\textbf{a}) Non-trivial geometric Zak phase landscape for ``split-step'' DTQW, obtained analytically; (\textbf{b}) non-trivial geometric Zak phase landscape for DTQW with non-commuting rotations, obtained by numeric integration \cite{PuentesCrystals17}.\label{f2}}
\end{figure}

A plot of the Zak phase is presented in Figure \ref{f2}a .

\subsection{Quantum Walk with Non-Commuting Rotations}

The DTQW with non-commuting rotations can be implemented via time multiplexing. The unitary step as described in the introduction results in $U(\theta,\phi)=TR_{x}(\phi)R_{y}(\theta)$.  The norms $n_i$ are of the following form
\begin{equation}
n_x=-\cos(k) a+\sin(k)b,
\end{equation}
\begin{equation}
 n_y=\cos(k) b+\sin(k) a,
\end{equation}
\begin{equation}
n_z=\cos(k) c-\sin(k) d,
\end{equation}
with
\begin{equation}
a=\sin(\phi)\cos(\theta),
\end{equation}
\begin{equation}
\lb{ang}
 b=\cos(\phi)\sin(\theta),
\end{equation}
\begin{equation}
 c=\sin(\phi)\sin(\theta),
\end{equation}
\begin{equation}
d=\cos(\phi)\cos(\theta),
\end{equation}
the  angular functions defined above. The numerator $N_1$ is given by
\begin{equation}\lb{n1}
N_1=n_x+in_y=- \exp(-i k)(a-i b).
\end{equation}

Then, the calculation of the Zak phase in terms of the eigenvectors ($|V_\pm\rangle$) for each band results in:
$$
Z=Z_{\pm} =i\int_{-\pi/2}^{\pi/2}dk \langle V_{\pm}|\partial_k V_{\pm}\rangle,
$$

By taking into account  (\ref{d}), the Zak phase is expressed as 
\begin{equation}
\label{eq:41}
Z_\pm=\int \frac{(a^2+b^2)dk}{D_\pm^2},
\end{equation}
\begin{eqnarray} \lb{d}
D_{\pm}&=&\sqrt{2n_x^2+ 2n_y^2+2n_z^2\mp 2n_z\sqrt{n_x^2+n_y^2+n_z^2}}\\
&=&\bigg(a^2+b^2+c^2 \cos^2(k)+d^2 \sin^2(k)-\sin(2k)cd\\
& &  \mp (\cos(k) c-\sin(k) d)\\
& & \times \sqrt{a^2+b^2+c^2 \cos^2(k)+d^2 \sin^2(k)-\sin(2k)cd}\bigg)^{\frac{1}{2}}.
\end{eqnarray}

We note that, in this example, the case $n_{z}=0$ is completely different than in the previous case, as~it returns a trivial Zak phase $Z=\pi$, since the k-dependence vanishes. We note that, for this system, the Zak phase landscape can be obtained by numerical integration. In particular, at the Dirac points indicated in Figure \ref{f1}, the Zak phase is not defined.
 
A plot of the Zak phase $\Phi_{Zak}$ is shown in Figure \ref{f2}b, for parameter values \mbox{$\theta_{1,2}=[-\pi, \pi]$} \mbox{and~$\phi=[-\pi, \pi]$}---(a) Zak phase for split-step quantum walk, given by the analytic expression $Z=\frac{\tan(\theta_2)}{\tan(\theta_1)}$; (b) Zak phase for quantum walk with non-commuting rotation obtained by numerical integration of  expression Equation \ref{eq:41}. 

It is well known that the Zak phase is a gauge dependent quantity---that is, it depends on the choice of origin of the unit cell \cite{ZakDemler}. For this reason, in general, it is not uniquely defined and it is not  a topological invariant. However, an invariant quantity can be defined in terms of the Zak phase \emph{difference} between two states ($|\psi^{\bf{1}} \rangle, |\psi^{\bf{2}} \rangle$) which differ on a geometric phase only. Generically, the Zak phase difference between two such states can be written as $\langle  \psi^{\bf{1}} |\psi^{\bf{2}}  \rangle=e^{i|\Phi_{Zak}^{\bf{1}} - \Phi_{Zak}^{\bf{2}}|}$. We stress that, by geometric invariance, we refer to properties that do not depend on the choice of origin of the Brillouin zone but only on relative distances between points in the Brillouin zone.

A simple experimental scheme to measure the Zak phase difference between states at any given step $N$ can be envisioned. For any choice of origin of the Brillouin zone, the system can be prepared by unitary evolution operators characterized by rotation parameters corresponding to either of the four adjacent Dirac points. A different geometric phase will be accumulated at each adjacent Dirac point. This phase difference can be measured by recombining the states, in the case of photons, by interfering with the states via a Mach-Zehnder interferometer. A suitable scheme for detection of the Zak phase difference in a photonic system is described in \cite{HolonomicWhite, PuentesCrystals17}. {Recently, a remarkable experimental implementation of Zak phase detection using orbital angular momentum was demonstrated by Cardano {\it et al.} \cite{Cardano17}.}

\subsection{Quantum Walk with Non-Local Coin Operation}

The entanglement between the coin and walker, induced by the coin and conditional translation operations, leads to the interference effects that makes the quantum walk to exhibit very different features from its classical counterpart. To explore the quantum nature of this phenomenon in its full extent, one should employ two particles, one embodying the coin, and the other one, the walker. Thus,~besides interference and entanglement, one would add the non-local feature to the process, which cannot be addressed in approaches employing two degrees of freedom of a single particle.

In Section~\ref{sec:QW_SLM}, we proposed an optical setup to prepare the walker-coin state in the $n$-th step of a 1D DTQW, using the polarization and transverse spatial modes of a single-photon field to encode the coin and the walker, respectively. Here, we generalize on this idea and propose a setup to prepare this same state, but now using two photons rather than one: the coin will be encoded in the polarization of the first photon and the walker will be encoded in the transverse modes of the second one. In this way, we expect to extend the range of phenomena that can be approached with the previous proposal to a regime where the coin operation and conditional translation are non-local. 

The referred walker-coin state for an unbiased coin operation (Equation~(\ref{eq:R})) and an initial state $|\psi_0\rangle=(a|H\rangle_1+b|V\rangle_1)|0\rangle_2$ with $|a|^2+|b|^2=1$, can be cast in the following form
\begin{equation}   \label{eq:non-local}
|\psi_n\rangle=(TR)^n|\psi_0\rangle=\frac{1}{\sqrt{2}}\left(|H\rangle_1|\varphi_{n}\rangle_2+|V\rangle_1|\xi_{n}\rangle_2\right),
\end{equation}
where 1 (2) labels the ``coin'' (``walker'') photon. The states $|\varphi_{n}\rangle$ and $|\xi_{n}\rangle$ are given by different superpositions of some of the transverse modes from $-n$ to $+n$, which depends on the initial coin state. For instance, if $n=4$ and the initial coin is $|H\rangle$, we can see from Equation~(\ref{eq:psi4}) that $|\varphi_4\rangle\propto|+4\rangle+\frac{3}{\sqrt{10}}|+2\rangle-\frac{1}{\sqrt{2}}(|0\rangle-|-2\rangle)$ and 
$|\xi_4\rangle\propto\frac{1}{\sqrt{10}}|+2\rangle+\frac{1}{\sqrt{2}}(|0\rangle-|-2\rangle)-|-4\rangle$. 

The two-photon state of Equation~(\ref{eq:non-local}) exhibits entanglement between polarization of photon 1 and spatial modes of photon 2; one cannot factorized it in states of these degrees of freedom individually. This so-called \emph{hybrid entanglement} \cite{Neves09} is, therefore, a useful resource for photonic implementations of quantum walk with non-local coin operations. 

\subsection{Applications via Spatial Light Modulators: Source of Non-Local Walker-Coin States Based on Two-Photon Hybrid Entanglement}

\begin{figure}[t]
\centering
\includegraphics[width=0.7\textwidth]{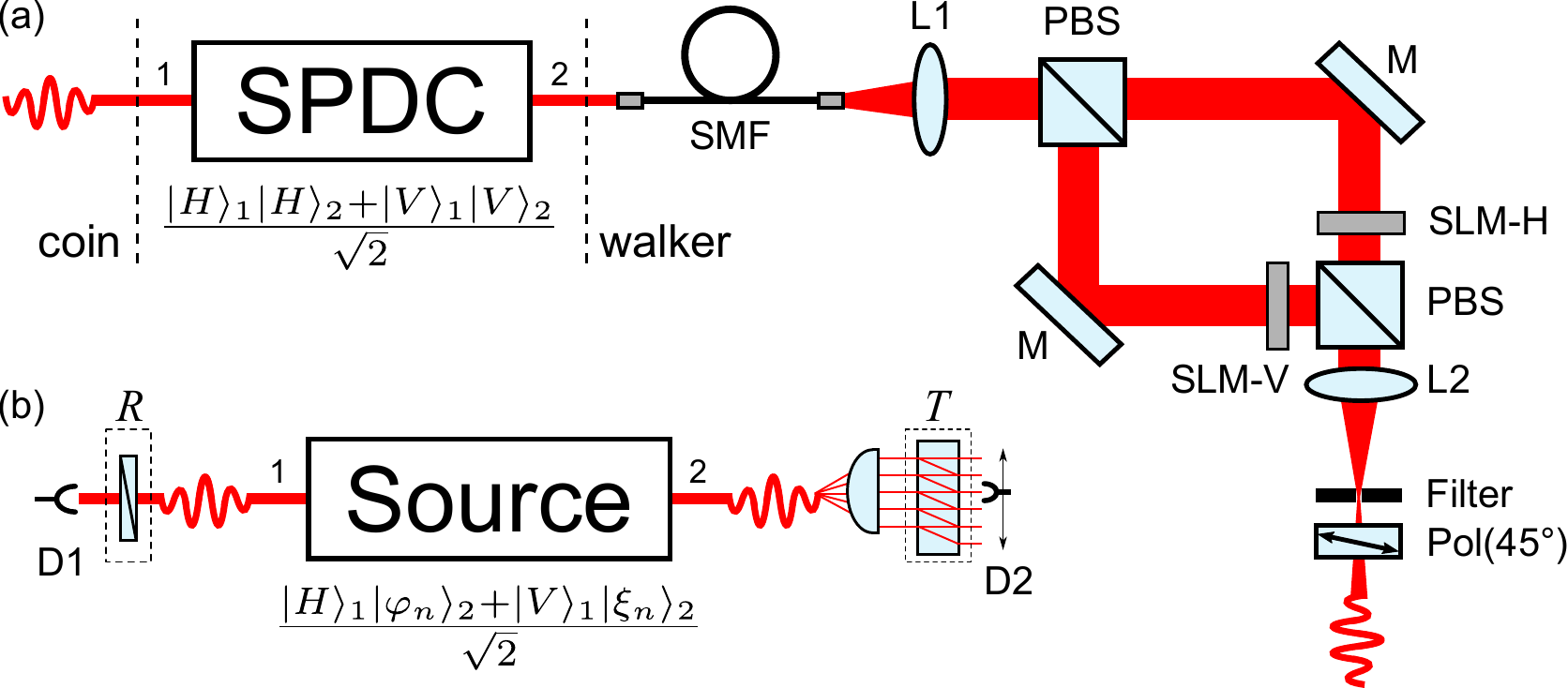}
\caption{\label{fig:HES} ({\bf a}) Schematic of a source for non-local walker-coin states (\ref{eq:non-local}) based on hybrid photonic entanglement. ({\bf b}) One step of a quantum walk with non-local coin operation using this source. The optical elements are the same of Figure~\ref{fig:SLM}d. See text for more details. }
\end{figure}

Figure~\ref{fig:HES}a illustrates the proposed  scheme for preparing hybrid entangled states given by Equation~(\ref{eq:non-local}): a~spontaneous parametric down-conversion (SPDC) source generates polarization-entangled photon pairs \cite{Kwiat99}. Assume that one of the photons, say 2, passes through an interference filter and a single mode fiber (SMF), which select a single frequency and a well defined gaussian mode for the pair, respectively. Thus, by properly configuring the source, one can generates maximally entangled states
\begin{equation}   \label{eq:entang_pol}
|\psi\rangle=\frac{1}{\sqrt{2}}\left(|H\rangle_1|H\rangle_2+|V\rangle_1|V\rangle_2\right)
\otimes|\gamma\rangle_1|\gamma\rangle_2,
\end{equation}
where $|\gamma\rangle_1|\gamma\rangle_2$ is the non-entangled spatial part of the two-photon state. Let us now look at the evolution of photon 2. After spatial filtering, its expanded transverse profile is collimated by a lens (L1) and it is sent to a polarization-based Mach-Zehnder interferometer. Both the input and output polarizing beam splitters (PBS) transmit (reflect) horizontally (vertically) polarized light. At each arm of the interferometer, there is a phase-only SLM with the working direction corresponding to the polarization in that arm. Each SLM will be addressed with a phase mask designed to prepare the walker's states $|\varphi_n\rangle$ and $|\xi_n\rangle$ of Equation~(\ref{eq:non-local}) (e.g., a mask like those shown in Figure~\ref{fig:SLM}c). The procedure for this is identical to the one briefly described in Section~\ref{sec:QW_SLM} [see the paragraph before Equation~(\ref{eq:chi_n})] and demonstrated in \cite{Prosser13}. At the output port of the interferometer the two arms are recombined, and after filtering the $+1$ diffraction order (where the spatial state is prepared), the two-photon state (\ref{eq:entang_pol}) is transformed into
\begin{equation}   
|\psi\rangle\rightarrow\frac{1}{\sqrt{2}}\left(|H\rangle_1|H\varphi_n\rangle_2+|V\rangle_1|V\xi_n\rangle_2\right),
\end{equation}
where the spatial component of photon 1, $|\gamma_1\rangle$, has been omitted, as it plays no role now. Finally, a linear polarizer (Pol) at $45^\circ$ projects the polarization of photon 2 in the state $|+\rangle=\frac{1}{\sqrt{2}}(|H\rangle+|V\rangle)$, thus generating the desired walker-coin state  (\ref{eq:non-local}), with the coin encoded in the polarization of photon 1, and the walker in the spatial modes of photon 2. It is important to stress that this interferometric approach can be also adopted for the single-photon scenario presented in Section~\ref{sec:QW_SLM}, with the difference that the projection in the polarization of photon 2 will not be applicable.

After preparing the state $|\psi_n\rangle$ given by  (\ref{eq:non-local}), one can submit it to a one-step evolution, $TR$, so that $|\psi_n\rangle\rightarrow|\psi_{n+1}\rangle$. For the single-photon case, the one-step optical module, described in Section~\ref{sec:QW_SLM}, is illustrated in Figure~\ref{fig:SLM}d. For the present case, this module must be dismembered with the coin operation $R$ (HWP at $22.5^\circ$) acting on photon 1, whereas photon 2 is subjected to a transverse-to-longitudinal mode conversion followed by the conditional translation $T$ provided by the cylindrical lens and beam-displacer, respectively. Figure~\ref{fig:HES}b illustrates this scenario. Here, the typical signature of the quantum walk, observed in the probability distribution of the walker's position, will be retrieved only in the coincidence basis. In this way, one guarantees that the global walker-coin state has undergone the evolution $TR$. Therefore, to determine $P_{n+1}(j)$ (Equation~(\ref{eq:P_n})), one must record the coincidence count rates between detectors $D1$ and $D2$ by scanning the later across the output modes of the beam-displacer.

\section{Conclusions}

In this review, we reported on different approaches to photonic discrete-time quantum walk (DTQW) architectures. Namely, implementations based on spatial- or temporal-mode multiplexing techniques, and implementations based on transverse photonic modes controlled by spatial light modulators (SLMs). We stress that while the number of steps ($n$) that can be implemented with multiplexed approaches is limited by the mode scaling of the multiplexed technique itself, i.e., $2n+1$ for spatial mode multiplexing and $2^n$ for temporal mode multiplexing, implementations using transverse modes enable simulation of an arbitrary step $n$, only limited by the resolution of the SLM itself.  We proposed different applications amenable to each DTQW architecture. In particular, we discussed the calculation of the Zak Phase, i.e.,  the Berry  phase across the Brillouin zone, for the case of the \emph{split-step} DTQW and for the case of DTQW with non-commuting rotations. We also proposed a novel application based on non-local coin operations using polarization-spatial two-photon hybrid entanglement, which can be readily implemented using parametric down-conversion and an interferometric arrangement of SLMs. 

\pagebreak
\subsection*{Acknowledgements}
\vspace{6pt}
The authors gratefully acknowledge Osvaldo Santillan, Marcos Saraceno and Mohammad Hafezi. 
G.P. gratefully acknowledges financial support from a PICT2014-1543 grant, a PICT2015-0710 grant, UBACyT PDE 2016, UBACYT PDE 2017  and Raices programme.  L.N. acknowledges financial support from FAPEMIG (APQ-00240-15) and INCT-IQ.

\subsection*{Author contributions}
Conceptualization G.P., Methodology, Software, Validation, and Formal Analysis L.N. and G.P., Original Draft Preparation G.P., Writing, Review and Editing, L.N. and G.P.. 



\end{document}